\documentclass[aps,twocolumn,prl]{revtex4}
\usepackage{epsfig}
\begin{document}

\title{Analytical Approach to Continuous and Intermittent Bottleneck Flows}
\author{Dirk Helbing, Anders Johansson}
\affiliation{Dresden University of Technology, Andreas-Schubert-Str. 23, 01062 Dresden, Germany}
\author{Joachim Mathiesen, Mogens H. Jensen, Alex Hansen}
\affiliation{Department of Physics, Norwegian University of Science and Technology,
7491 Trondheim, Norway, and 
The Niels Bohr Institute, Blegdamsvej 17, 2100 Copenhagen, Denmark}
\date{\today}

\begin{abstract}
We propose a many-particle-inspired theory for granular outflows from a hopper and 
for the escape dynamics through a bottleneck based on a continuity equation 
in polar coordinates. If the inflow is below the maximum outflow, we find an asymptotic
stationary solution. If the inflow is above this value, we observe
queue formation, which can be described 
by a shock wave equation. We also address the
experimental observation of intermittent outflows, taking 
into account the lack of space in the merging zone by a minimum function and
coordination problems by a stochastic variable. This results in avalanches of
different sizes even if friction, force networks, inelastic collapse, or delay-induced
stop-and-go waves are not assumed. Our intermittent flows result from a
random alternation between particle propagation and gap propagation. 
Erratic flows in congested merging zones of vehicle traffic may be 
explained in a similar way.
\end{abstract}

\maketitle

Driven granular media display a rich spectrum of pattern formation phenomena.
This includes collective oscillating states, convection patterns, 
the spontaneous segregation of different granular
materials, and the formation of avalanches due to self-organized criticality \cite{review}.  
Here, we will focus on jamming and clogging phenomena \cite{granclog}
related to arching \cite{arching}, 
and intermittent outflows through hoppers \cite{inter1,Clement}.
Similar phenomena are
known from dense pedestrian crowds \cite{pedopus}. 
The escape dynamics of individuals from a room has been intensively studied,
showing that in crowd stampedes, 
rooms are emptied in an irregular, strongly intermittent fashion
\cite{panic}. This effect has been discovered in 
simulations performed with the social and the 
centrifugal force model \cite{panic,ped}, with cellular automata 
and lattice gas automata \cite{ped1}, and in a meanfield model \cite{naga2}. 
It has also been experimentally confirmed \cite{pedopus,densecrowds}. 
However, analytical models of escape dynamics and granular
bottleneck flows are lacking.
\par
In this Letter we will formulate such a model. Our goal is to gain a better
understanding of (i) the resulting density profiles and (ii) the irregular outflows at bottlenecks.
This includes not only the distribution of the avalanche sizes in the outflow from a bottleneck.
We will also offer a possible explanation of the long-standing problem of 
perturbations forming in merging zones of freeway traffic flows
\cite{kerner,daganzo}, which are characterized by erratic, forward or backward moving shock waves \cite{kerner}. 
It is believed that these can trigger stop-and-go waves in traffic flows \cite{kerner,analytic}. 
Similar findings have been made in over-crowded pedestrian flows \cite{densecrowds}
and expected  for merging flows in urban traffic and production networks.
\par
In all these cases, the competition 
of too many entities for little space leads to coordination problems. 
We are therefore looking for a minimal, common model capturing this feature. Hence,
we will first abstract from specific system features such as the non-Newtonian character of real granular 
flows, non-slip boundary conditions, dissipative interactions, 
or force networks in quasi-static granular flows \cite{Kadanoff,Copper}, 
and discuss extensions later.
This will allow us to show that intermittent flows are caused
{\it even without} mechanisms like dissipative collapse \cite{Kadanoff}, 
large spatio-temporal fluctuations due to force networks \cite{Copper}, 
or delay-induced instabilities (as in traffic flows). These may 
magnify the effect \cite{NOTE}.
\par
As pedestrian evacuation has been successfully described by driven granular particle 
models, where a single particle represents an individual pedestrian, we will formulate a common model 
for escaping pedestrians and gravity-driven outflows 
from vertical, two-dimensional hoppers. Due to the conservation of the 
particle number, we will describe the aggregate, two-dimensional particle flow by the continuity 
equation for the particle density $\rho$ as a function of space and time. Both, the shape of a funnel and the
semicircular shape of a waiting crowd suggest to write this equation in polar coordinates. 
Assuming no significant dependence on the polar angles $\theta$ and $\varphi$ for the moment, 
we obtain
$\partial \rho/\partial t + (1/r) \partial (r \rho v)/\partial r = 0$ 
(generalizing this to a 2d treatment later). 
Here, $t$ denotes the time, $r\ge 0$ the distance from the bottleneck (exit) and 
$v\le 0$ the velocity component in radial direction. 
The above continuity equation can be rewritten as 
\begin{equation}
 \frac{\partial \rho}{\partial t} + \frac{\partial (\rho v)}{\partial r} = - \frac{\rho v}{r} \, ,
\label{cont}
\end{equation}
where the term on the right-hand side reflects a merging effect similar to
an on-ramp or lane closure term in a model of freeway traffic. 
By use of logarithmic derivatives, the above equation can be rewritten as
$\partial \ln \rho(r,t)/\partial t = -  v(r,t)  \partial \ln [r \rho(r,t) v(r,t)]/\partial r$.
For the stationary case with $\partial \ln \rho/\partial t = 0$ it follows from 
$\partial \ln (r \rho v)/\partial r = 0$ that the overall flow
$f \pi r \rho(r) v(r)$ through any cross section at distance $r$ 
is constant: 
\begin{equation}
 f \pi r \rho(r) v(r) = f  \pi r q(r) =: - Q_0  = \mbox{const.} 
\label{C}
\end{equation} 
$q(r) = \rho(r) v(r)$ is the particle flow through a cross section 
of unit length. $f=1$ corresponds to the half circumference $\pi r$ of a circle of radius $r$, 
while $f < 1$ allows one to treat hoppers with an opening angle smaller than 180 degrees.
(The walls should be steeper than the angle of repose.)
$Q_0 \ge 0$ is the stationary overall particle flow. 
\par
\begin{figure}[htbp]
\includegraphics[width=.49\textwidth]{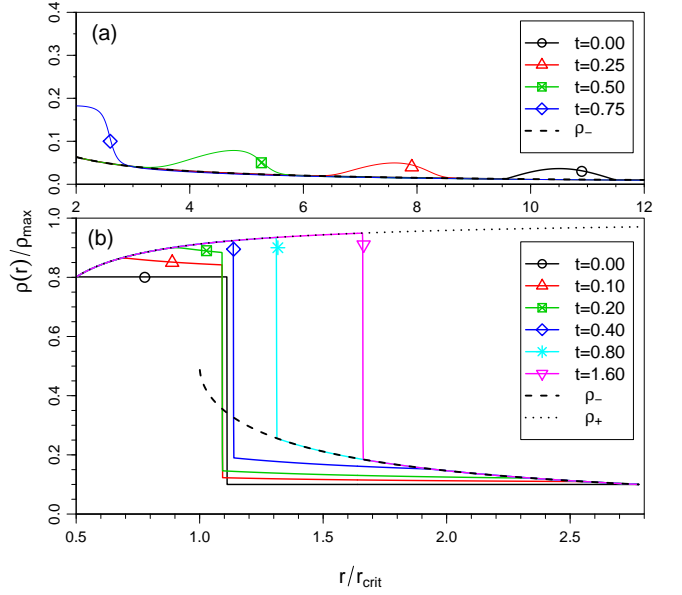}
\caption[]{(Color online)Density profiles at different times, when (a) the inflow is low 
and the initial density profile has a hump, 
(b) the inflow exceeds the maximum outflow and the initial density profile is a step function 
(shock wave). The simulation results have been obtained by solving the continuity equation
with the Godunov scheme, assuming 
$\rho(12r_{\rm crit},t)= 0.01$ and floating boundary conditions at 
$r=r_0 = 2r_{\rm crit} := 2r_{\rm crit}(Q_{\rm in})$
in case (a), but $\rho(2.8r_{\rm crit},t)= 0.1$ and 
$Q_{\rm out} = 2r_0 q_{\rm max}$ (corresponding to the maximum
outflow) with $r_0 = 0.5r_{\rm crit}$ in case (b).
Note that the asymptotic density profile is $\rho_-(r,Q_{\rm in})$ in free flow and
$\rho_+(r,Q_{\rm out})$ in jammed flow.}\label{Fig1}
\end{figure}

To facilitate the derivation of analytical results, 
we will assume the linear velocity-density relationship
\begin{equation}
 v(r) = V\big(\rho(r)\big) = -v_0 \left( 1 - \rho/\rho_{\rm max} \right) \le 0 \, .
\label{velocity}
\end{equation}
$v_0$ means the maximum (``free'') particle speed and 
$\rho_{\rm max}$ the maximum particle density.
Eqs. (\ref{C}) 
and (\ref{velocity}) give the quadratic equation
$\rho(r)v_0[1-\rho(r)/\rho_{\rm max}] = Q_0/(f \pi r)$ in $\rho$. 
With $r_{\rm crit}(Q_0) := Q_0/(f\pi  q_{\rm max})$ it implies 
\begin{equation}
\rho_{\pm}(r,Q_0) 
 = \frac{\rho_{\rm max}}{2} \left( 1 \pm \sqrt{ 1 - \frac{r_{\rm crit}(Q_0)}{r}}\right) \, , 
\label{dens}
\end{equation}
where $q_{\rm max} = v_0 \rho_{\rm max}/4$ is the maximum flow. 
In free flow with $d|\rho V(\rho)|/d\rho \ge 0$, the density profile is determined
by the upstream boundary condition, i.e. $Q_0$ is given by the overall inflow
$Q_{\rm in}$. Under congested conditions ($d|\rho V(\rho)|/d\rho < 0$), 
$Q_0$ is given by the overall outflow 
$Q_{\rm out} = \min(Q_{\rm in}, 2r_0 q_{\rm max},f\pi r_0  q_{\rm max})$, i.e. 
the minimum of the overall inflow $Q_{\rm in}$ and the maximum possible overall outflow. 
The stationary case requires $Q_{\rm in}=Q_{\rm out}$ and
a non-negative discriminant in Eq.~(\ref{dens}). 
This calls for $r_0 \ge r_{\rm crit}(Q_{\rm in})$, i.e. large outlets (see Fig.~\ref{Fig1}a). Then, $\rho(r,t)$ converges
to a stationary free flow with the density profile $\rho_-(r,Q_{\rm in}) \le \rho_{\rm max}/2$. 
The density profile for other velocity-density relationships than (\ref{velocity}) can be obtained numerically.
Smooth perturbations like the humps in 
Fig. \ref{Fig1}a propagate forward at the speed $V(\rho)+\rho dV(\rho)/d\rho
= -v_0 (1-2\rho/\rho_{\rm max})$ \cite{Whitham}, compactify close to the outlet and leave the system.
\par
If the overall inflow exceeds the overall outflow ($Q_{\rm in}>Q_{\rm out}$), particles are jammed 
behind the outlet (Fig. \ref{Fig1}b). 
The initial density profile $\rho(r,0)$ again approaches $\rho_-(r,Q_{\rm in})< \rho_{\rm max}/2$ in the free-flow
regime at large distances $r$, but converges to $\rho_+(r,Q_{\rm out})>\rho_{\rm max}/2$ in the congested
regime upstream of the outlet. The congestion front moves forward until the jam density 
$\rho_+(r,Q_{\rm out})$ is reached. 
Then, the particles queue up and the shock front at location $R(t)$ 
moves backward at the speed
\begin{equation}
 \frac{dR}{dt} = - \frac{Q_{\rm in} - Q_{\rm out}}{f\pi R(t) [\rho_-(R,Q_{\rm in}) - \rho_+(R,Q_{\rm out})]} 
\end{equation}
according to the shock wave equation \cite{Whitham} (see Fig. \ref{Fig1}b). 
Hence, we find the free-flow density profile
$\rho(r,t) \approx \rho_-(r,Q_{\rm in})$ for $r> R(t)$, while for $r< R(t)$ we have the 
congested density profile $\rho(r,t) \approx \rho_+(r,Q_{\rm out})$.
\par
\begin{figure}
\includegraphics[width=.5\textwidth]{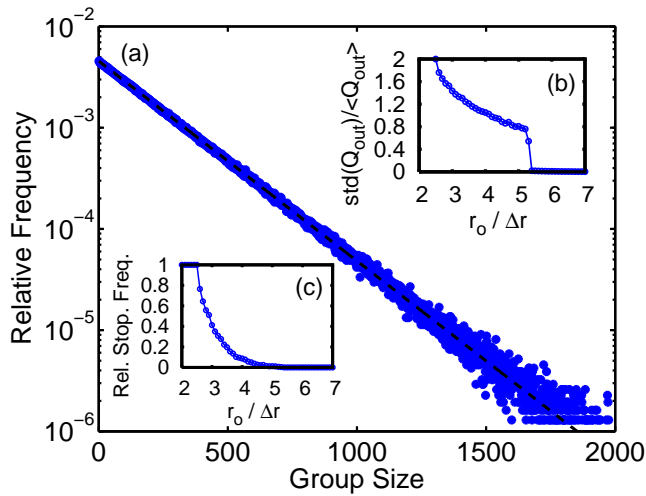}
\caption[]{(Color online)(a) In agreement with an experiment for granular
outflows from a two-dimensional hopper \cite{Clement},  
our simulation model generates exponentially distributed 
avalanche sizes when frictionless particles with coordination problems are jamming at
a bottleneck (i.e. theory and experiment show a straight line in a log-linear plot). 
(b) The standard deviation of the outflow, divided by the average outflow shows
3 regimes: no outflow for $r_0/\Delta r < 1/\gamma = 2.5$, smooth outflows for
large outlets, and intermittent flow in between. (c) The relative proportion of time
steps $\Delta t$ with a stopped outflow confirms this picture.
Our results are quite insensitive to the selected parameters.
For illustration, we chose $\beta=3$, $\gamma =2/5$, $\epsilon = 0.01$, 
$Q_{\rm in} = 4/\Delta t$, $v_0 = \Delta r/\Delta t$, $\rho_{\rm max} = 1/(\Delta r)^2$
and, in (a), $r_0 = 5\,\Delta r$ (jammed conditions).}\label{Fig2}
\end{figure}

 
This applies to cases of continuous outflows, which are observed 
for large enough openings \cite{ped,granclog}  or small enough pedestrian 
velocities $v_0$ \cite{panic}. 
However, if the desired velocity $v_0$ of pedestrians is
high, their maximum density $\rho_{\rm max}$ goes up and intermittent outflows
are observed \cite{panic,densecrowds}. 
This intermittent behavior (see Fig.~\ref{Fig2}) reminds of driven granular media \cite{inter1} 
and shall be modeled now. 
For this, let us subdivide the particle bulk into shells of thickness $\Delta r$ (for example, the
particle diameter $d$ or multiples of it). Within each shell of area 
$A(r) \approx f \pi r \, \Delta r$, we assume a constant average
density $\rho(r,t) = N(r,t)/A(r)$, where $N(r,t)$ denotes the number of particles in 
the shell of radius $r$ at time $t$. Furthermore, we assume that particles move from one
shell to the next with velocity $v_0$, if they find a suitable gap, otherwise, they will
stay. The maximum number of particles available to move into the shell of radius $r$ is
$\rho(r+\Delta r,t) A(r+\Delta r)$, while the maximum number of available gaps
in shell $r$ is $\rho_{\rm max} A(r) [1 - \rho(r,t)/\rho_{\rm max}]$, because $\rho_{\rm max}A(r)$
is the maximum number of particles in the shell of radius $r$ and 
$q(r,t) = 1 - \rho(r,t)/\rho_{\rm max}$ represents the fraction of free space. 
Finally, we assume that $\xi_r^\pm  q(r,t)$ denotes the probability to find a suitable gap in front of a particle
allowing it to move, where $\xi_r^\pm$ are random numbers specified in each time step
with $0\le \xi_r^\pm \le 1$ and $\xi_{r-\Delta r}^+ = \xi_r^-$ (in order to guarantee particle conservation).
Then, the number of inflowing particles within the time interval $\Delta t = \Delta r/v_0$
is $N_{\rm in}(r,t) = \xi_r^+ q(r,t) \min [ A(r+\Delta r)\rho(r+\Delta r, t), A(r)\rho_{\rm max} ]$,
while the number of outflowing particles is $N_{\rm out}(r,t)
= \xi_r^- q(r-\Delta r,t) \min [ A(r)\rho(r, t), A(r-\Delta r)\rho_{\rm max} ]$. From the
balance equation $N(r,t+\Delta t)= N(r,t) + N_{\rm in}(r,t) - N_{\rm out}(r,t)$ and
$\rho(r,t) = N(r,t)/A(r)$ we get
\begin{eqnarray}
 \rho(r,t+\Delta t) 
 &=& \rho(r,t) + \xi_r^+ \left( 1 - \frac{\rho(r,t)}{\rho_{\rm max}} \right) \nonumber \\
 &\times& \min \left[ \left(1+\frac{\Delta r}{r}\right) \rho(r+\Delta r, t), \rho_{\rm max} \right] \nonumber \\
 &-& \xi_r^-  \left( 1 - \frac{\rho(r-\Delta r,t)}{\rho_{\rm max}} \right) \nonumber \\
 &\times& \min \left[ \rho(r, t), \left(1-\frac{\Delta r}{r}\right) \rho_{\rm max} \right] \, . \qquad
\label{min}
\end{eqnarray}
Finally note that the half circle of radius $r_0$ around the exit is treated analogously to the shells, 
but we have to replace the area $A(0)$ by $A_0 = \pi r_0{}^2/2$ and $-N_{\rm out}(0,t)$ by
$-2r_0 \rho_0(t)v_0\, \Delta t$ (i.e. the exit width $2r_0$ times the flow $q_0(t)=\rho_0(t) v_0$, 
if pedestrians can leave with maximum velocity $v_0$ into the uncongested space behind the exit). 
The resulting equation for the density $\rho_0(t)$ in the last (sub-)area before passing the bottleneck is
\begin{eqnarray}
 \rho_0(t+\Delta t) &=& \rho_0(t) + \xi_0^+(t) \left( 1 - \frac{\rho_0(t)}{\rho_{\rm max}} \right) \nonumber \\ 
 &\times &\min \left[ \frac{2\Delta r}{r_0} \rho(r_0,t),\rho_{\rm max} \right] - \frac{4\Delta r}{\pi r_0} \rho_0(t) \, . \qquad  
\label{eleven}
\end{eqnarray}
The minimum function in Eq.~(\ref{min}) 
delineates the merging-related lack of space and outflow capacity.
A similar situation and minimum function occurs in merging flows in urban street and production
networks. 
With $\Delta r = v_0\, \Delta t$, $\rho = \rho(r,t)$, $\xi_r = (\xi_r^++\xi_r^-)/2$, $\zeta_r = 
(\xi_r^+ -\xi_r^-)(1-\rho/\rho_{\rm max})/\Delta t$ 
and for $\rho(r,t)\le (1-\Delta r/r) \rho_{\rm max}$ 
we find the following equation in the limit $\Delta t, \Delta r \rightarrow 0$: 
\begin{equation}
 \frac{\partial \rho}{\partial t} = v_0\left(\xi_r^+ - 2 \xi_r \frac{\rho}{\rho_{\rm max}} \right) 
 \frac{\partial \rho}{\partial r}
+ \frac{\rho v_0 \xi_r^+}{r} \left( 1 - \frac{\rho}{\rho_{\rm max}} \right) 
+ \zeta_r \rho \, . 
\end{equation}
With the linear velocity-density relation (\ref{velocity}), this exactly corresponds
to the previous continuity equation (\ref{cont}), if 
$\xi_r^\pm = 1$, as for small enough densities (see below). 
Fluctuations $\xi_r^\pm <1$, however,
allow one to describe a dynamics in which less particles than possible are successful 
in finding a gap in the next shell, because of coordination problems. The random 
variable $\xi_r^\pm$ reflects that the microscopic spatial configuration of the particles matters.  
When the second terms in the minimum functions of Eq.~(\ref{min})
apply, the dynamics is given by the equation
\begin{equation}
 \frac{\partial \rho}{\partial t} + v_0 \xi_r^- \frac{\partial \rho}{\partial r} 
 = \frac{v_0\xi_r^-}{r} [\rho_{\rm max} - \rho(r,t)] + \zeta_r \rho_{\rm max} \, . 
\end{equation} 
After averaging over the noise terms $\xi_r^\pm$, representing the average of $\xi_r^-$ by
$\overline{\xi}$, defining the the gap density $\hat{\rho}(r,t) = \rho_{\rm max} - \rho(r,t)$,
and introducing $\widehat{V} = v_0\overline{\xi}$, this turns into
a continuity equation for gap propagation:
\begin{equation}
 \frac{\partial \hat{\rho}}{\partial t} + \frac{\partial (\hat{\rho}\widehat{V})}{\partial r}
 = - \frac{\hat{\rho}\widehat{V}(t)}{r} \, . 
\end{equation}
Note that gaps propagate with velocity $\widehat{V}>0$, i.e. in opposite direction to the particles.
\par
We expect that a switching between gap propagation and particle propagation 
by the minimum function can account for the intermittent outflows of dense granular flows. Triggered by
the randomness of the variable $\xi_r^\pm$, the switching mechanism can produce particle avalanches of different sizes.
The fluctuations $\xi_r^\pm$ and their average value $\overline{\xi}$ can be adjusted to 
experimental or suitable microsimulation results, e.g. to reflect the spatio-temporal fluctuations due to
granular force networks. Here, we have instead simulated Eq.~(\ref{min}) with binomially distributed values of
$\xi_r^\pm$, i.e. $\xi_r^\pm = k/N$ with $P(k) = \left( 
\begin{array}{c}
N \\
k
\end{array} \right)
p^k (1-p)^{N-k}$. $P(k)$ is the probability that $k\in \{0,1,\dots, N\}$ of 
$N=N(r,t)$ particles successfully manage 
to move forward, where $p = \overline{\xi}$ is the probability of a particle not to be obstructed. 
We have used the phenomenological specification 
\begin{equation}
 p(\rho,r) = 
\left\{ \!1 + \!\left[ \frac{r}{\Delta r} \!\left(\frac{\rho_{\rm max}}{\rho} - 1\right)^\beta 
\!\!\!+ \epsilon \left(\gamma - \frac{\Delta r}{r} \right)
\right]^{-1} \! \right\}^{\!-1} 
\end{equation}
if $p>0$, otherwise $p:=0$. ($\beta$, $\gamma$, and $\epsilon$ are non-negative fit parameters.) 
This ensures that $p(\rho,r)$ becomes 1 for $\rho \rightarrow 0$ or $r\rightarrow \infty$ 
and $\max[0,\epsilon (\gamma - \Delta r/r )]$ for $\rho \rightarrow
\rho_{\rm max}$. That is, we have complete clogging, if $r_0/\Delta r < 1/\gamma \approx 5/2$, 
which reflects
arching if the outlet $2r_0$ is too small (see Fig.~\ref{Fig2}). 
Otherwise, if the density is low or the bottleneck far away, 
we have $p=1$ corresponding to particles moving at the speed $v=V(\rho)$. 
In queued areas with $\rho\approx \rho_{\rm max}$, gaps propagate upstream with velocity
$\widehat{V} = v_0 \epsilon (\gamma - \Delta r/r )$.
\par
We have presented a shockwave approach to determine the spatio-temporal density profile
in granular bottleneck flows and evacuation scenarios. Generalized to two dimensions allowing
to consider boundary conditions, friction, etc., there is a free-flow regime characterized by 
forward motion according to 
$\partial \rho/\partial t + \partial (\rho V)/\partial x = - [\rho V/w(x)] dw(x)/dx$,
where $(dw/dx)/w = (\partial V_{\perp}/\partial x_\perp)/V$ replaces $1/r$ and describes the relative 
change of locally available width $w$, i.e. the bottleneck effect. $x$ is the coordinate {\it in} flow 
direction and $V>0$ the corresponding speed, while $\perp$ represents the perpendicular direction.
If $w(x) \rho > [w(x) + \Delta x\, dw/dx] \rho_{\rm max}$,
the density after the next step of length $\Delta x$ (where $\Delta x$ is the mean free path or a fit parameter) 
would exceed the maximum possible density $\rho_{\rm max}$. Therefore, if the
``gap density'' $\hat{\rho}(x,t) := \rho_{\rm max} 
- \rho(x,t)$ falls below the value $- \rho_{\rm max} \Delta x\, dw/dx $, we have instead
the equation $\partial \hat{\rho}/\partial t + \partial (\hat{\rho}\widehat{V})/\partial x
 = -[\hat{\rho} \widehat{V}/w(x)] dw(x)/dx$ 
for gap propagation, where $\widehat{V}<0$ is the backward propagation speed. 
Hence, at bottlenecks we have alternating phases of forward pedestrian motion with
speed $V$ and of upstream dissolving pedestrian jams with average speed $\widehat{V}$, 
where $\widehat{V}$ and $V$ may fluctuate in space and time. 
\par
These formulas are useful for the appropriate 
dimensioning of exits in order to avoid critical situations in cases of emergency evacuation of people. If the
bottleneck is too small (and the desired speed $v_0$ of pedestrians too high), one may find intermittent flows
close to and behind the bottleneck (Fig. \ref{Fig2} b, c). 
These are due to the fact that too many ``particles'' are competing
for a confined space. Obviously, not all particles can successfully progress, 
when there are mutual obstructions. This ``coordination problem'' has been reflected by a 
fluctuation factor $\xi_r^\pm$, the mean value 
$p(\rho,w/(dw/dx))$ of which drops significantly below 1 if $\rho(r,t)/\rho_{\rm max} > 1+ \Delta r \, (dw/dx)/w(x)$,
i.e. if not all particles fit into the reduced space when progressed by a distance $\Delta x$. 
\par
The resulting dynamics is related to a stop-and-go phenomenon: In the high-density
jam, the velocity is zero, as the particles cannot move. However, jam resolution at the exit
causes an upstream moving shock wave, in front of which the density is low. Therefore, 
particles at the jam front can leave the jam. In fact, if the density in front of the 
jam is small enough, there is a forward motion of particles filling the low-density area.
Altogether, we will have alternating phases of jam resolution and gap filling 
processes close to the exit, which lead to alternating propagation directions of the 
jam front. This may also explain the observed alternation
in the propagation direction of perturbations in freeway traffic flows \cite{kerner}.\\
Support by the DFG (He2789/7-1) is acknowledged.


\begin{thebibliography}{20}
\bibitem{NOTE} and also change its statistics (e.g. stick-slip friction would cause self-organized
criticality with power-law distributed avalanche sizes \cite{inter1}).
\end{thebibliography}
\end{document}